\documentclass{PoS}

\title{Relevant momentum components of gluons for confinement and chiral symmetry breaking}

\ShortTitle{Relevant momentum components of gluons for confinement and $\chi$SB}

\author{\speaker{Arata~Yamamoto} and Hideo~Suganuma\\
Department of Physics, Faculty of Science, Kyoto University, Sakyo, Kyoto 606-8502, Japan\\
E-mail: \email{a-yamamoto@ruby.scphys.kyoto-u.ac.jp}}

\abstract{
We investigate which momentum components of gluons induce color confinement and spontaneous chiral symmetry breaking in lattice QCD.
For this purpose, we formulate a lattice framework to introduce the momentum cutoff of the gluon field.
Using this framework, we calculate the quark-antiquark potential, the color flux tube, the chiral condensate, and the Dirac spectrum.
Our results suggest that confinement and chiral symmetry breaking are induced by somehow different momentum components of gluons.
}

\FullConference{The XXVIII International Symposium on Lattice Field Theory\\
June 14-19, 2010\\
Villasimius, Sardinia, Italy}

\begin{document}

\section{Introduction}
The strong interaction is mediated by gluons.
The interaction strength of gluons, i.e., the QCD running coupling constant, depends on the energy scale.
For this reason, the behaviors of QCD phenomena are different in the different energy scales.
The analysis for the role of momentum components of gluons is important for understanding the mechanism of QCD phenomena.

In this study, our goal is to reveal which momentum components of gluons induce each QCD phenomenon.
In particular, we focus on two of the most significant properties of QCD, color confinement and spontaneous chiral symmetry breaking.
The connection between confinement and chiral symmetry breaking is often discussed in QCD.
In some effective theories, the energy scales of these two phenomena are different at zero temperature \cite{Ma84}.
On the other hand, deconfinement and chiral symmetry restoring phase transitions occur at the same temperature \cite{Ko83}.
We nonperturbatively explore this connection in the viewpoint of the gluon momentum.
We calculate the quark-antiquark potential \cite{Ya08,Ya09}, the color flux tube \cite{YaS1}, the chiral condensate \cite{Ya10}, and the Dirac spectrum \cite{YaS2} in the quenched lattice QCD simulations at zero temperature.

In general, since the gauge transformation is nonlocal in momentum space, the momentum component of the gluon field is not gauge invariant.
We must fix the gauge in the numerical simulation, and its result would depend on the adopted gauge choice.
Nevertheless, this kind of analysis enables us to intuitively understand the roles of gluons for QCD phenomena.
In this paper, we show the numerical results in the Landau gauge.

\section{Momentum cutoff}
We introduce the lattice framework to remove some momentum components of the gluon field \cite{Ya08}.
The framework is formulated as the following five steps.

{\it Step 1}. The SU(3) link variable $U_{\mu}(x)$ is generated by Monte Carlo simulation.
The link variable is fixed with a certain gauge.

{\it Step 2}. The momentum-space link variable ${\tilde U}_{\mu}(p)$ is obtained by the Fourier transformation, as
\begin{eqnarray}
{\tilde U}_{\mu}(p)=\frac{1}{V}\sum_{x} U_{\mu}(x)\exp(i \sum_{\nu} p_\nu x_\nu ),
\end{eqnarray}
where $V$ is the lattice volume.

\begin{figure}[t]
\begin{center}
\includegraphics[scale=0.4]{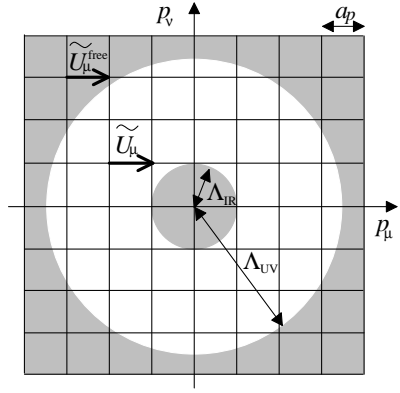}
\caption{\label{fig1}
The schematic figure of the UV cutoff $\Lambda_{\rm UV}$ and  the IR cutoff $\Lambda_{\rm IR}$.
}
\end{center}
\end{figure}

{\it Step 3}. Some components of ${\tilde U}_{\mu}(p)$  are removed by a momentum cutoff.
In the cutoff region, the momentum component is replaced by the free link variable
\begin{equation}
{\tilde U}^{\rm free}_{\mu}(p)=\frac{1}{V}\sum_x 1 \cdot \exp(i {\textstyle \sum_\nu} p_\nu x_\nu)=\delta_{p0}.
\end{equation}
For example, the ultraviolet (UV) cutoff $\Lambda_{\rm UV}$ is
\begin{equation}
{\tilde U}_{\mu}^{\Lambda}(p)= \Bigg\{
\begin{array}{cc}
{\tilde U}_{\mu}(p) & (\sqrt{p^2} \le \Lambda_{\rm UV})\\
0 & (\sqrt{p^2} > \Lambda_{\rm UV}),
\end{array}
\end{equation}
and the infrared (IR) cutoff $\Lambda_{\rm IR}$ is
\begin{equation}
{\tilde U}_{\mu}^{\Lambda}(p)= \Bigg\{
\begin{array}{cc}
\delta_{p0} & (\sqrt{p^2} < \Lambda_{\rm IR})\\
{\tilde U}_{\mu}(p) & (\sqrt{p^2} \ge \Lambda_{\rm IR}).
\end{array}
\end{equation}
The schematic figure is depicted in Fig.~\ref{fig1}.

{\it Step 4}. The coordinate-space link variable is obtained by the inverse Fourier transformation as
\begin{eqnarray}
U'_{\mu}(x)=\sum_{p} {\tilde U}_{\mu}^{\Lambda}(p)\exp (-i \sum_{\nu} p_\nu x_\nu ).
\end{eqnarray}
Since $U'_{\mu}(x)$ is not an SU(3) matrix in general, $U'_{\mu}(x)$ must be projected onto an SU(3) element $U^{\Lambda}_{\mu}(x)$.
The projection is realized by maximizing the quantity
\begin{eqnarray}
{\rm ReTr}[\{ U^{\Lambda}_{\mu}(x) \}^{\dagger}U'_{\mu}(x)].
\end{eqnarray}

{\it Step 5}. The expectation value of an operator $O$ is computed by using this link variable $U^{\Lambda}_{\mu}(x)$ instead of $U_{\mu}(x)$, i.e., $\langle O[U^\Lambda]\rangle$ instead of $\langle O[U]\rangle$.

Using this framework, we analyze how the physical quantity is changed by the momentum cutoff at fixed gauge.
From the resultant change, we can nonperturbatively investigate the relation between the physical quantity and the momentum components of gluons.
We expect that this framework is broadly applicable to the analysis of QCD phenomena.

\section{Color confinement}
First, we apply this framework to the static quark-antiquark potential \cite{Ya09}.
The quark-antiquark potential is expressed as $V(R)=\sigma R -A/R +C$ with the interquark distance $R$.
In short range, it is dominated by the perturbative Coulomb potential.
In long range, it is dominated by the nonperturbative linear confinement potential.

\begin{figure}[t]
\begin{center}
\includegraphics[scale=1]{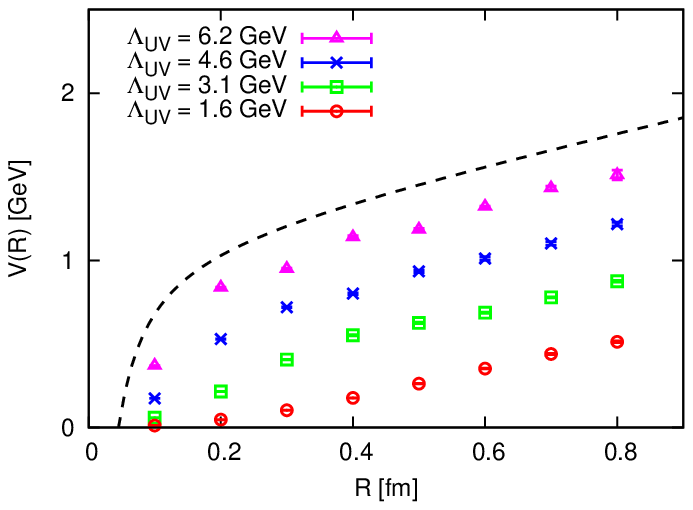}
\includegraphics[scale=1]{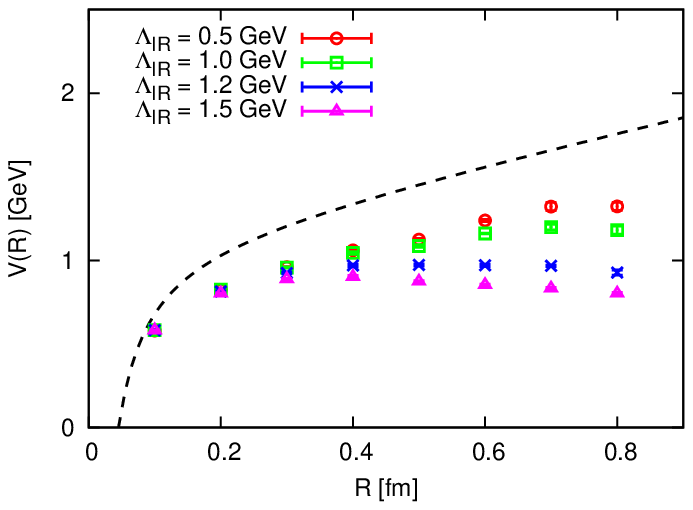}
\caption{\label{fig2}
The quark-antiquark potential $V(R)$ with the UV cutoff $\Lambda_{\rm UV}$ (left) and  the IR cutoff $\Lambda_{\rm IR}$ (right).
The broken curve is the original quark-antiquark potential.
}
\end{center}
\end{figure}

\begin{figure}[t]
\begin{center}
\includegraphics[scale=1]{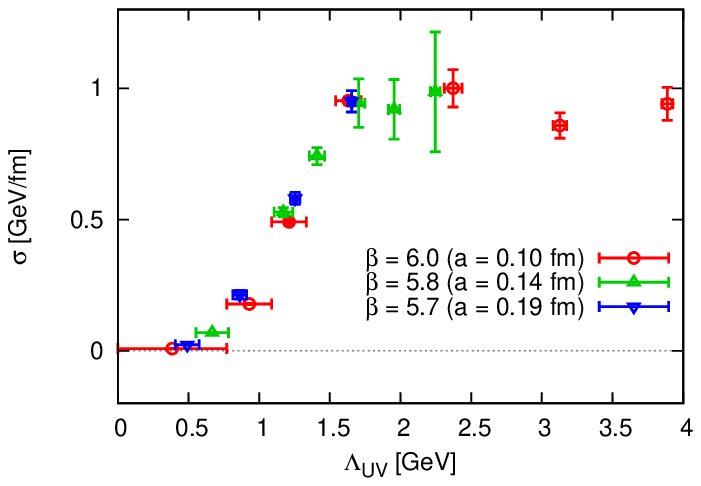}
\includegraphics[scale=1]{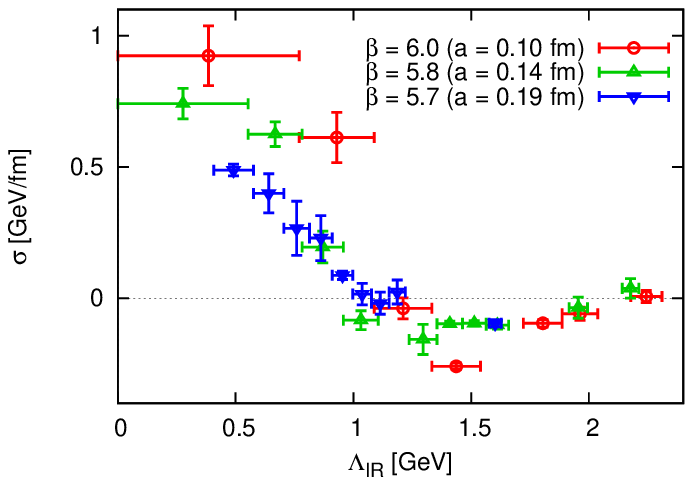}
\caption{\label{fig3}
The string tension $\sigma$ with the UV cutoff $\Lambda_{\rm UV}$ (left) and  the IR cutoff $\Lambda_{\rm IR}$ (right).
The results with different lattice spacings $a$ are shown.
}
\end{center}
\end{figure}

We show the quark-antiquark potential with the momentum cutoff in Fig.\ref{fig2}.
When the high-momentum component is removed by the UV cutoff, the Coulomb potential disappears and the quark-antiquark potential becomes a linear potential.
When the low-momentum component is removed by the IR cutoff, the linear confinement potential disappears, and the quark-antiquark potential becomes the perturbative Coulomb potential.

For a more quantitative argument, we estimate the string tension $\sigma$ by fitting the potential with a linear function $\sigma R + {\rm const.}$ in $0.3 \ {\rm fm}< R < 0.9$ fm.
The results with different lattice spacings are summarized in Fig.~\ref{fig3}.
The original value of $\sigma$ is about 0.89 GeV/fm.
Both in the cases of UV and IR cutoffs, the string tension is drastically changed in the low-momentum region below about 1.5 GeV.
On the other hand, the string tension is almost unchanged in the high-momentum region above about 1.5 GeV.
Therefore, we conclude that color confinement is induced by the low-momentum gluon below about 1.5 GeV.

\begin{figure}[t]
\begin{center}
\includegraphics[scale=1.2]{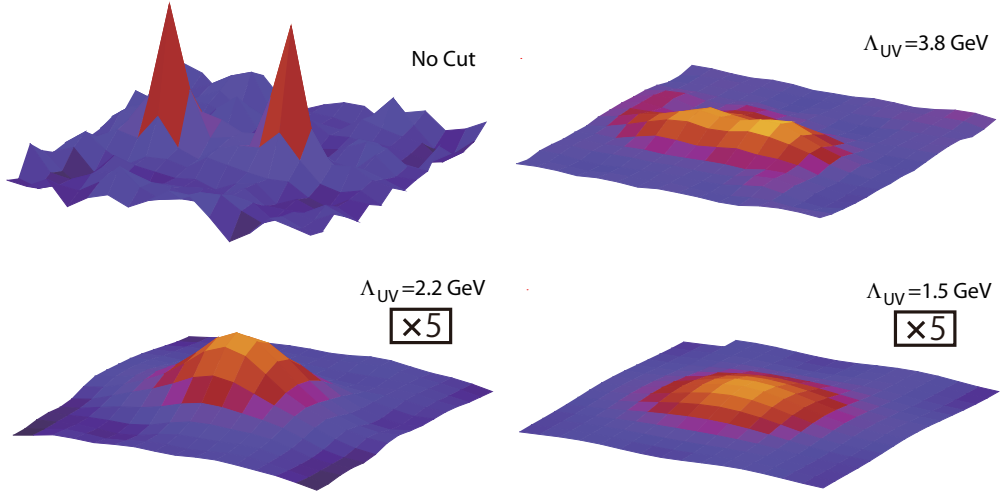}
\caption{\label{fig4}
The action density distribution around a static quark-antiquark pair.
The interquark distance is $R=0.6$ fm.
``$\times 5$'' means that the action density is enlarged by a factor of five compared to other data.
}
\end{center}
\end{figure}

Based on this conclusion, we analyze the color flux tube \cite{YaS1}.
In hadrons, gluons form the color flux tube between the confined quarks.
The formation of the color flux tube is believed to be essential for color confinement.
In lattice QCD, one can observe the color flux tube by calculating the spatial action density distribution around the Wilson loop \cite{So87,Gi90,Ba95,Ha96,Pe97}.
The resultant action density distribution is shown in Fig.~\ref{fig4} (No Cut).
Note that the statistical fluctuation is rather large.
There are two singular peaks which correspond to quark self energies.
The color flux tube exists between these two peaks, which is, however, too small compared to perturbative contributions.

Since color confinement is induced only by the low-momentum gluon below about 1.5 GeV, we can expect that the color flux tube is extracted by removing the high-momentum gluon above 1.5 GeV.
Here, for a technical reason, we use the three-momentum cutoff, which gives almost the same behavior as the four-momentum cutoff \cite{YaS1}.
In Fig.~\ref{fig4}, we display the action density distribution with the UV cutoff.
When the high-momentum gluon is removed, the perturbative self-energy peaks are strongly suppressed.
In addition, as a by-product, the statistical fluctuation is also strongly suppressed.
The action density distribution with $\Lambda_{\rm UV}=1.5$ GeV purely corresponds to the color flux tube.

\section{Spontaneous chiral symmetry breaking}
Next, we analyze spontaneous chiral symmetry breaking.
Although chiral symmetry is, of course, a symmetry of the quark field, spontaneous chiral symmetry breaking is induced by the nonperturbative gluon dynamics.

\begin{figure}[t]
\begin{center}
\includegraphics[scale=1.1]{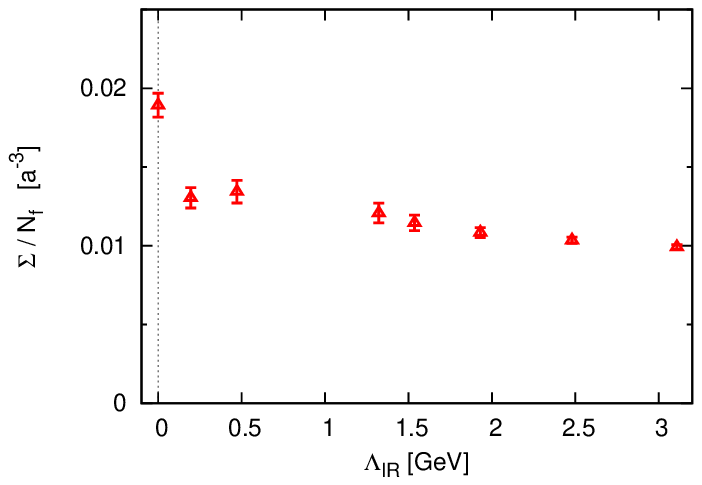}
\includegraphics[scale=1.1]{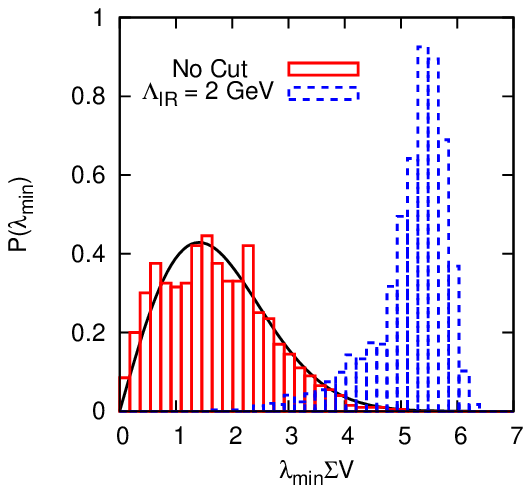}
\caption{\label{fig5}
The chiral condensate $\Sigma=| \langle \bar{q}q \rangle |$ with the IR cutoff $\Lambda_{\rm IR}$ (left).
The probability distribution of the smallest Dirac eigenvalue $\lambda_{\rm min}$ (right).
The solid curve is the prediction of chiral random matrix theory.
}
\end{center}
\end{figure}

An order parameter of spontaneous chiral symmetry breaking is the chiral condensate $\langle \bar{q}q \rangle$ in the chiral limit.
In Fig.~\ref{fig5}, we show $\Sigma=| \langle \bar{q}q \rangle |$ with the IR cutoff in quenched lattice QCD \cite{Ya10}.
In this calculation, we use the staggered fermion with the mass $ma=0.01$, where the corresponding pion mass is about 500 MeV.
Around $\Lambda_{\rm IR}=0$, the chiral condensate suddenly gets small.
This jump around $\Lambda_{\rm IR}=0$ is caused by removing the zero-momentum link variable ${\tilde U}_{\mu}(0)$.
This suggests that the zero-momentum link variable, which is the deep-infrared gluon in the continuum, has a large contribution to the chiral condensate.
In $\Lambda_{\rm IR}>0$, the chiral condensate gradually decreases.
Even in $\Lambda_{\rm IR}>1.5$ GeV, the chiral condensate continues to decrease.
Also in the case of other quark masses and in the extrapolated chiral limit, the qualitative behavior is similar.

In this quenched calculation, the chiral condensate is induced by the broad momentum region.
This is different from the case of color confinement.
This fact suggests that color confinement and spontaneous chiral symmetry breaking is induced by the different momentum components of gluons.
For more conclusive statement, we have to take into account many systematics, especially, the dynamical quark effects in the case of realistic QCD.

The chiral condensate is related to the Dirac spectrum through the Banks-Casher relation \cite{Ba80}.
Thus, the gradual change of the chiral condensate is considered to originate from the nonlinear relation between the Dirac eigenvalue and the gluon momentum.
This relation can be directly analyzed by calculating the Dirac spectrum \cite{YaS2}.

Another interesting property of the low-lying Dirac spectrum is that it is described by chiral random matrix theory \cite{Sh93,Go99,Da99}.
Chiral random matrix theory describes the universal behavior of disordered systems.
In Fig.~\ref{fig5}, we show the smallest eigenvalue distribution $P(\lambda_{\rm min})$ in the trivial topological sector.
The histograms are the lattice data of the improved staggered Dirac operator.
The solid curve is the prediction of chiral random matrix theory, $P(\lambda_{\rm min}) = z_{\rm min}/2 \cdot \exp (-z_{\rm min}^2/4)$ with $z_{\rm min}=\lambda_{\rm min}\Sigma V$ \cite{Fo93}.
In the original lattice QCD (No Cut), the lattice data is well reproduced by chiral random matrix theory.
In $\Lambda_{\rm IR}=2$ GeV, the lattice data deviates from the prediction of chiral random matrix theory.
Therefore, the low-momentum gluon induces the strong-interacting and disordered nature which is necessary for the validity of chiral random matrix theory.

\section*{Acknowledgments}
A.~Y.~and H.~S.~are supported by a Grant-in-Aid for Scientific Research [(C) No.~20$\cdot$363 and (C) No.~19540287] in Japan.
This work was in part based on the MILC collaboration's public lattice gauge theory code (http://physics.utah.edu/\~{}detar/milc.html).
The lattice QCD simulations were carried out on Altix3700 BX2 and SX8 at YITP in Kyoto University, and SX8 at Osaka University.


\begin{thebibliography}{99}
\bibitem{Ma84} A.~Manohar and H.~Georgi, Nucl. Phys. B {\bf 234}, 189 (1984).
\bibitem{Ko83} J.~B.~Kogut, M.~Stone, H.~W.~Wyld, W.~R.~Gibbs, J.~Shigemitsu, S.~H.~Shenker, and D.~K.~Sinclair, Phys. Rev. Lett. {\bf 50}, 393 (1983).
\bibitem{Ya08} A.~Yamamoto and H.~Suganuma, Phys. Rev. Lett. {\bf 101}, 241601 (2008).
\bibitem{Ya09} A.~Yamamoto and H.~Suganuma, Phys. Rev. D {\bf 79}, 054504 (2009).
\bibitem{YaS1} A.~Yamamoto, Phys. Lett. B {\bf 688}, 345 (2010).
\bibitem{Ya10} A.~Yamamoto and H.~Suganuma, Phys. Rev. D {\bf 81}, 014506 (2010).
\bibitem{YaS2} A.~Yamamoto, arXiv:1005.2241 [hep-lat].
\bibitem{So87} R.~Sommer, Nucl. Phys. B {\bf 291}, 673 (1987).
\bibitem{Gi90} A.~Di~Giacomo, M.~Maggiore, and S.~Olejnik, Nucl. Phys. B {\bf 347}, 441 (1990).
\bibitem{Ba95} G.~S.~Bali, C.~Schlichter, and K.~Schilling, Phys. Rev. D {\bf 51}, 5165 (1995).
\bibitem{Ha96} R.~W.~Haymaker, V.~Singh, Y.~C.~Peng, and J.~Wosiek, Phys. Rev. D {\bf 53}, 389 (1996).
\bibitem{Pe97} P.~Pennanen, A.~M.~Green, and C.~Michael, Phys. Rev. D {\bf 56}, 3903 (1997).
\bibitem{Ba80} T.~Banks and A.~Casher, Nucl. Phys. B {\bf 169}, 103 (1980).
\bibitem{Sh93} E.~V.~Shuryak and J.~J.~M.~Verbaarschot, Nucl. Phys. A {\bf 560}, 306 (1993).
\bibitem{Go99} M.~G\"{o}ckeler, H.~Hehl, P.~E.~L.~Rakow, A.~Sch\"{a}fer, and T.~Wettig, Phys. Rev. D {\bf 59}, 094503 (1999).
\bibitem{Da99} P.~H.~Damgaard, U.~M.~Heller, R.~Niclasen, and K.~Rummukainen, Phys. Rev. D {\bf 61}, 014501 (1999).
\bibitem{Fo93} P.~J.~Forrester, Nucl. Phys. B {\bf 402}, 709 (1993).
\end{thebibliography}
\end{document}